\documentclass{emulateapj}
\usepackage{graphicx}

\def\gsim{\lower 2pt \hbox{$\, \buildrel {\scriptstyle >}\over
{\scriptstyle \sim}\,$}}
\def\lsim{\lower 2pt \hbox{$\, \buildrel {\scriptstyle <}\over
{\scriptstyle \sim}\,$}}

\def\civ{C~{\scriptsize IV}}

\def\nv{N~{\sc v}}

\def\ovi{O~{\sc vi}}

\slugcomment{ }

\shorttitle{Variable High-Ion Emission from LMC X-3}
\shortauthors{Song et al.}

\begin{document}

\title{Variable \ovi\ and \nv\ Emission from the X-ray Binary LMC X-3: \\ 
Heating of the Black Hole Companion\altaffilmark{1}}

\altaffiltext{1}{Based on observations with (1) the NASA/ESA {\it
Hubble Space Telescope}, obtained at the Space Telescope Science
Institute, which is operated by the Association of Universities for
Research in Astronomy, Inc., under NASA contract NAS 5-26555, (2)
the NASA-CNES/ESA {\it Far Ultraviolet Spectroscopic Explorer}
mission, operated by Johns Hopkins University, supported by NASA
contract NAS 5-32985, and (3) the 6.5 meter Magellan
Telescopes located at Las Campanas Observatory, Chile.}

\author{Limin Song,\altaffilmark{2} Todd M. Tripp,\altaffilmark{2}
  Q. Daniel Wang,\altaffilmark{2} Yangsen Yao,\altaffilmark{3} Wei Cui,\altaffilmark{4} 
Yongquan Xue,\altaffilmark{4,5}
Jerome A. Orosz,\altaffilmark{6} Danny Steeghs,\altaffilmark{7} 
James F. Steiner,\altaffilmark{8} Manuel A. P. Torres,\altaffilmark{8} Jeffrey E. McClintock
\altaffilmark{8}}
\altaffiltext{2}{Department of Astronomy, University of Massachusetts, Amherst, MA 01003, USA}
\altaffiltext{3}{Center for Astrophysics and Space Astronomy, University of Colorado, Boulder, CO 80309, USA}
\altaffiltext{4}{Department of Physics, Purdue University, West Lafayette, IN 47907, USA}
\altaffiltext{5}{Department of Astronomy and Astrophysics, 525 Davey Lab,
Pennsylvania State University, University Park, PA 16802, USA}
\altaffiltext{6}{Department of Astronomy, San Diego State University,
5500 Campanile Drive, San Diego, CA 92182-1221, USA}
\altaffiltext{7}{Department of Physics, University of Warwick, Coventry, CV4 7AL, UK
and Harvard-Smithsonian Center for Astrophysics, 60 Garden Street, Cambridge, MA
02138, USA}
\altaffiltext{8}{Harvard-Smithsonian Center for Astrophysics, 60 Garden Street,
Cambridge, MA 02138, USA}

\begin{abstract}
Based on high-resolution ultraviolet spectroscopy obtained with the
{\it Far Ultraviolet Spectroscopic Explorer (FUSE)} and the Cosmic
Origins Spectrograph, we present new detections of \ovi\ and
\nv\ emission from the black-hole X-ray binary (XRB) system LMC
X-3. We also update the ephemeris of the XRB using recent radial
velocity measurements obtained with the echelle spectrograph on the
Magellan-Clay telescope.  We observe significant velocity variability
of the UV emission, and we find that the \ovi\ and \nv\ emission
velocities follow the optical velocity curve of the XRB.  Moreover,
the \ovi\ and \nv\ intensities regularly decrease between binary
phase = 0.5 and 1.0, which suggests that the source of the UV emission
is increasingly occulted as the B star in the XRB moves from superior
to inferior conjunction.  These trends suggest that illumination of
the B-star atmosphere by the intense X-ray emission from the accreting
black hole creates a hot spot on one side of the B star, and this hot
spot is the origin of the \ovi\ and \nv\ emission.  However, the
velocity semiamplitude of the ultraviolet emission, $K_{\rm UV}
\approx$ 180 km s$^{-1}$, is lower than the optical semiamplitude;
this difference could be due to rotation of the B star. Comparison of
the {\it FUSE} observations taken in November 2001 and April 2004
shows a significant change in the \ovi\ emission characteristics: in
the 2001 data, the \ovi\ region shows both broad and narrow emission
features, while in 2004 only the narrow \ovi\ emission is clearly
present. {\it Rossi X-ray Timing Explorer} data show that the XRB was
in a high/soft state in the Nov. 2001 epoch but was in a transitional
state in April 2004, so the shape of the X-ray spectrum might change
the properties of the region illuminated on the B star and thus change
the broad vs. narrow characteristics of the UV emission. If our
hypothesis about the origin of the highly ionized emission is correct, then
careful analysis of the emission occultation could, in principle,
constrain the inclination of the XRB and the mass of the black hole.
\end{abstract}


\section{Introduction}
How do black holes affect and interact with their surroundings?  This
is one of the highest priority questions of current astrophysics.
Typically, the angular momentum of accreting material drives the
formation of a disk around the black hole.  The hot inner accretion
disk is a powerful source of high-energy X-ray radiation, since
immense gravity can cause substantial heating of matter spiraling in
toward the black hole. In general, the X-rays emerging from the inner
disk and corona can illuminate and ionize surrounding material, and in the case
of black holes in binary star systems, the high-energy emission can
affect matter on the surface of the companion star, in the stellar
wind, in the accretion flows, or the outer disk.  A significant amount
of the X-ray radiation could be reprocessed into ultraviolet/optical
photons. Such reprocessing might account for the continuum and/or some
of the observed emission features in the UV/optical spectra of
black-hole systems (although UV/optical continuum emission can also
emanate directly from the accretion disk).  For this reason, the study
of UV spectra of black hole systems can provide valuable insights
about the physics of accreting black holes and the impact they have on
their surroundings. The analysis of the variability of the UV
radiation from a binary star system as a function of orbital phase
can determine the location of the region producing the UV photons, the
physical conditions existing in the region, the mechanism producing
the radiation, and kinematic parameters of the binary itself.

LMC X-3, which was first discovered by \citet{leong71},
is an excellent target for this type of
study. It is a bright X-ray binary (XRB) system in the Large
Magellanic Cloud composed of a B3 V star (V $\sim 16.7-17.5$) and a
central black-hole (Cowley et al. 1983, hereafter C83). It is one of the few
black-hole candidates that are persistently luminous in X-rays
and also bright enough to be observed with high-resolution ultraviolet
spectrographs with good signal-to-noise (S/N). The black hole is
believed to be undergoing accretion from its B-star companion via
Roche lobe overflow with an orbital period of 1.7 days (C83; Soria et al. 2001).
Spectroscopic observations of the B star indicate a large radial velocity
semiamplitude, $K_B=235 \pm 11 \ km \ s^{-1}$
(C83). Although LMC X-3 has been intensively studied in the
X-ray band, UV spectroscopic observations of this target are more
sparse.  The first \textit{Far Ultraviolet Spectroscopic Explorer}
({\it FUSE}) observation of LMC X-3 was made by Hutchings et al. (2003, hereafter H03)
in November 2001. During the 24 ks exposure, LMC X-3 was in its
brightest X-ray phase. These observations revealed a broad emission
feature in the vicinity of the \ovi\ $\lambda \lambda$1031.9,1037.6
doublet, which Hutchings et al. argue is due to the blend of the
1031.9 and 1037.6 \AA\ lines (see their Figure 4).  From this
emission, they measured a shift in the \ovi\ emission
velocity between two binary phases of about 100-150 km
s$^{-1}$. Assuming the \ovi\ emission arises in the inner parts of the
black hole's accretion disk, they conclude that the minimum mass of
the B star and the black hole are 13 and 15 $M_{\odot}$,
respectively. However, their measurement of the velocity variation of
\ovi\ emission with binary phase suffers from some
uncertainties. First, the ephemeris they used to determine the phase
is from more than 20 years ago. Over such a long time baseline, the
errors contained in the ephemeris could be accumulated, resulting in
phases that are off by $\approx$0.1 at the $1 \sigma$ level. Another
source of uncertainty is the limited phase coverage of their
observations. Due to the modest S/N of their {\it FUSE} data,
Hutchings et al. summed their {\it FUSE} spectra into only two phase
bins centered at phases = 0.53 and 0.70, which only loosely constrains
$K_{bh}$, the velocity semiamplitude of the black hole. Overall, these
limitations preclude any firm conclusions about the nature and
implications of the \ovi\ emission.

An updated ephemeris and expanded UV spectroscopy with better orbital
coverage are needed to better constrain the implications of the UV
emission lines and the characteristics of this XRB system. In this
paper, we revisit LMC X-3 with a new ephemeris and new UV
observations.  We update the ephemeris of LMC X-3 based on
high-resolution optical spectroscopy recently obtained with the 6.5m
Magellan-Clay telescope. We also report new ultraviolet observations
of LMC X-3 from two high-resolution instruments. First, we present
temporal monitoring of the \ovi\ emission using {\it FUSE} data with a
much longer time baseline that provides more than three quarters
of the LMC X-3 orbit.  With the extended {\it FUSE} data, we detect narrow \ovi\ 
emission from the binary system and measure the variations in velocity
and intensity of the emission as a function of orbital phase. Second,
we complement the \ovi\ analysis with new observations of the \nv\
$\lambda \lambda$1238.8,1242.8 doublet obtained with the Cosmic
Origins Spectrograph (COS) on board the {\it Hubble Space Telescope
(HST)}.  The \nv\ emission is detected at high significance in the COS
data and provides corroborating evidence of the velocity and intensity
variations of the highly ionized emission.  During the time of the {\it
FUSE} observations in 2004, LMC X-3 was also observed
quasi-simultaneously in the X-ray bandpass with {\it Chandra} and the
{\it Rossi X-ray Timing Experiment (RXTE)}; see \citet{wang05} for
full details. While our emphasis in this paper is on the UV emission
lines, we also briefly comment on the affiliated X-ray
observations. With the combination of this information, we investigate
the following questions: (1) Does the \ovi\ emission originate in the
accretion disk, the ionized stellar wind, the illuminated surface of
the B star, or a hot spot (e.g., from the spot where overflowing Roche
lobe material impacts the accretion disk)? (2) Does the \nv\ doublet
emission arise in the same region as the \ovi\ emission, i.e., are the
\nv\ and \ovi\ features physically associated?

\section{Observations}
\subsection{Optical Spectroscopy\label{sec:opt_obs}}

Fifty three spectra of the optical counterpart 
(Warren \& Penfold 1975) of the X-ray binary system 
were obtained on the nights of 2005 January
20--24, 2007 December 20--21, and 2008 February 27 -- 2008 March
1 using the Magellan Inamori Kyocera Echelle (MIKE) spectrograph
\citep{bernstein03} on the 6.5 m Magellan-Clay telescope at Las
Campanas Observatory. 
The instrument was used in the
standard dual-beam mode with a $1\farcs 0\times5\farcs 0$ slit
and the $2\times 2$ binning mode.
With a few exceptions, the typical exposure time was 1800 s.
We only focus on spectra from the blue arm, which has a wavelength
coverage of 3430--5140 \AA\ and resolving power of $R=33,000$.
The radial velocities were determined from these
new spectra using the {\sc fxcor} task within IRAF and then fitted to
a circular orbit model. The typical uncertainties of the radial 
velocities are 3.5 km s$^{-1}$. 

\subsection{Ultraviolet Spectroscopy}

\subsubsection{{\it FUSE} Observations}
Following the initial observations of \citet{hutchings03}, LMC X-3 was
observed again with {\it FUSE} in April 2004, and during the
latter campaign quasi-simultaneous observations were recorded with
{\it Chandra} and {\it RXTE} \citep{wang05}.  The 2004 {\it FUSE}
observations provided a total exposure time of 86 ks ($\approx 4
\times$ longer than the original {\it FUSE} observations obtained 
in H03). LMC X-3 is a variable source,
and in April 2004 the average UV flux of the target turned out to be
$\approx 4 \times$ fainter in the relevant spectral regions compared
to the UV flux when \citet{hutchings03} observed it in November
2001. Consequently, the S/N ratios are comparable in the {\it FUSE}
data obtained in 2001 and 2004. However, the addition of the later
data set significantly improves the orbital coverage of the \ovi\
data. Both the 2001 and 2004 {\it FUSE} data were reduced with
CALFUSE, version 3.0 \citep{dixon07}. It is important to note that the
{\it FUSE} SiC channels suffer from scattered solar light problems in
the 2004 data, and in this paper we only use the data from the LiF
channels \citep[details regarding the design and performance of {\it
FUSE} can be found in][]{moos00,moos02}.

\begin{figure*}
\centering
    \includegraphics[width=16.0cm, angle=0]{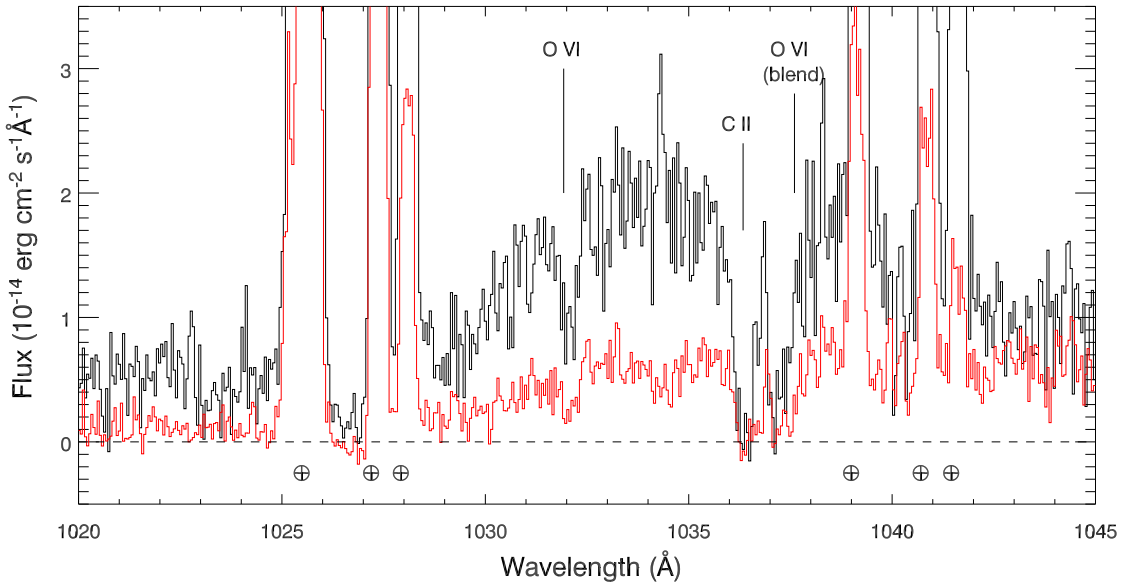}
\caption{Comparison of the {\it FUSE} observations of LMC X-3 obtained
in 2001 (black histogram, see H03) and in 2004
\citep[red histogram, see][]{wang05} in the vicinity of the \ovi\
$\lambda$1031.9,1037.6 doublet.  In this figure, all data from
the November 2001 observing campaign have been accumulated into a
single spectrum regardless of the phase of the binary during the
individual exposures, and likewise all of the April 2004 observations
have been coadded to form a single spectrum. The \ovi\ doublet emission, whose 
velocities and intensities vary with the binary phase as presented
in Figures~\ref{vE063} and \ref{velstack}, are smeared out in the combined spectra.
Both spectra are binned to 7.5 km s$^{-1}$
pixels. Airglow emission lines from the Earth's outer atmosphere are
marked ($\oplus$), and absorption lines due to \ovi\ and C~{\sc ii} ($\lambda$1036.3)
are also labeled. The weaker \ovi\ absorption line at $\lambda$1037.6
is blended with Galactic absorption due to C~{\scriptsize II}$^*$ ($\lambda$1037.0) and 
$H_2$ ($\lambda$1037.2). It might also
be blended with C II at the velocity of the LMC itself. \label{OVIcmp}}
\end{figure*}

In Figure~\ref{OVIcmp}, we compare the {\it FUSE} observations of LMC
X-3 obtained in 2001 (black histogram) and in 2004 (red histogram) in
the wavelength range near the \ovi\ doublet (in this figure, all of
the data obtained in 2001 are accumulated into a single spectrum, and
likewise all of the 2004 data are coadded). 
Bright airglow emission
lines from the Earth's atmosphere are marked ($\oplus$), and
well-detected absorption lines from the \ovi\ doublet and {\sc C~ii}
$\lambda$1036.3 near $v$(heliocentric) $\approx$ 0 km s$^{-1}$ are
also labeled \citep[analysis of the absorption lines in the {\it FUSE}
spectrum of LMC X-3 is reported in][]{wang05}.  The substantial
variation in the UV flux is immediately obvious, but we note that the
flux variation cannot be described with a simple scale factor applied
to the entire spectrum.  For example, the continuum is roughly two times 
brighter at $\lambda \approx$ 1045 \AA\ but is roughly $4\times$ brighter near the
putative \ovi\ emission at $\lambda \approx$ 1034 \AA .  It is
difficult to draw conclusions about the long-term UV flux variability
with observations at only two epochs, but it appears that while the
\ovi\ emission intensity decreased along with the continuum in the
2004 observation, there is not a strict correlation.  We will show
below that narrow \ovi\ emission lines are detected in the 2004 data,
but it is possible that the {\it broad} \ovi\ emission was not present
in the later observation for some reason.  The shape of the X-ray
spectrum of LMC X-3 also changed between 2001 and 2004 (see \S
\ref{sec:xrayobs}), so it is possible that the changes of the broad
\ovi\ feature are related to changes in the X-ray flux.

\subsubsection{COS Observations}
There are some limitations of the spectroscopic UV dataset from {\it
FUSE} presented in the previous section.  First, LMC
X-3 is a challenging target for {\it FUSE} because the UV flux is
relatively low, so the data can be noisy.  Also, in the {\it
FUSE} wavelength range, emission lines from the Earth's atmosphere
are a potential source of confusion (see below).  Moreover, {\it FUSE}
is no longer operating, so we cannot contemplate follow-up {\it FUSE}
spectroscopy. Fortunately, all of these problems can be overcome with
COS \citep{green03}, the new spectrograph on {\it HST}, with some
caveats.  COS has limited sensitivity in the 912 $< \lambda <$
1130 \AA\ range \citep{mccandliss10}, but this region can only be
observed with COS at low spectral resolution ($R \approx$ 2000). COS
could observe LMC X-3 with higher resolution ($R \approx 20,000$) at
$\lambda > 1130$ \AA, but at the expense of missing the OVI doublet.
However, other highly ionized species are accessible in the $\lambda > 1130$
\AA\ range (e.g., \nv\ $\lambda \lambda$1238.8, 1242.8 or
\civ\ $\lambda \lambda$1548.2,1550.8).

\begin{figure*}
\centering
    \includegraphics[width=8.0cm, angle=0]{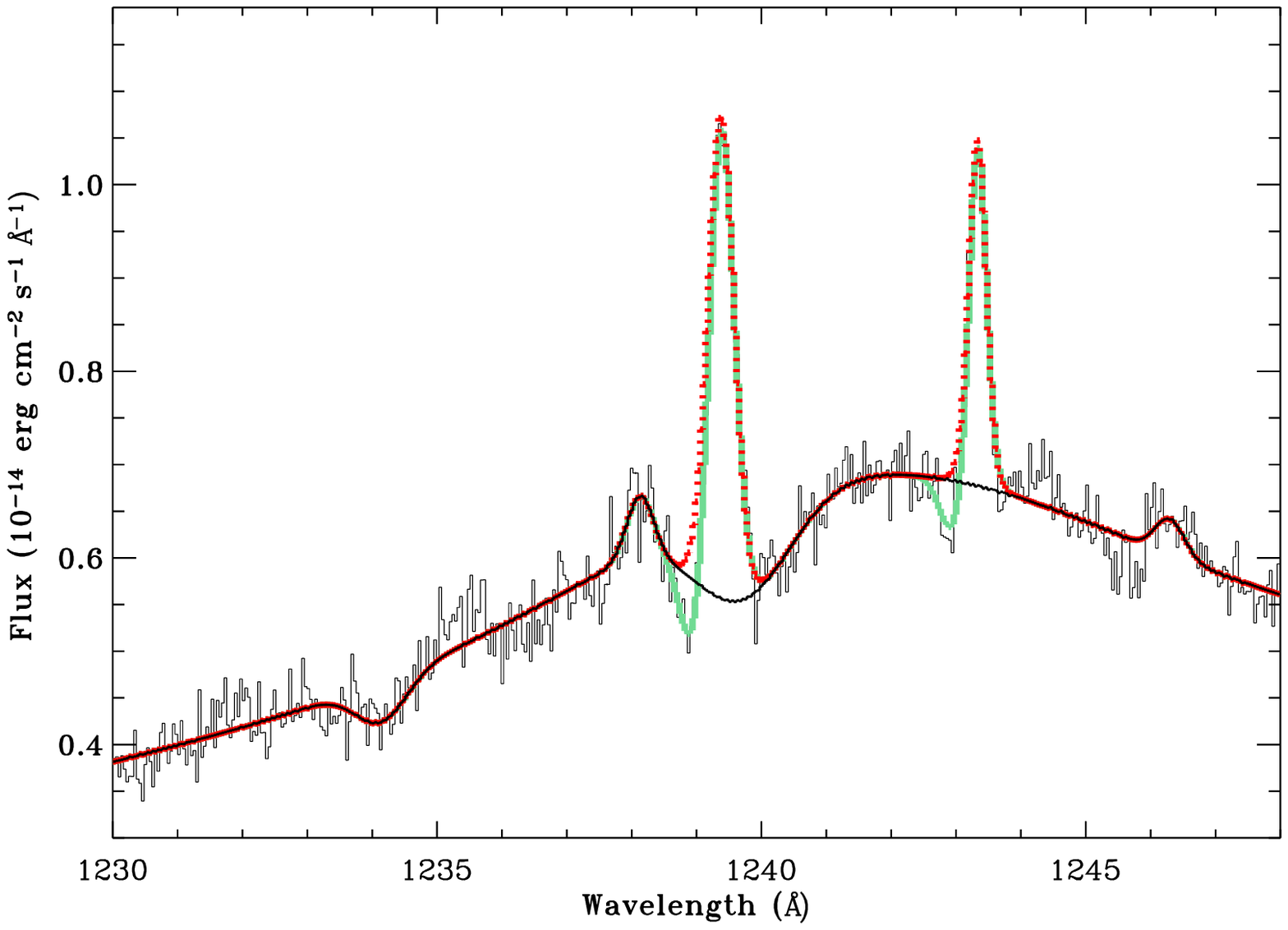}
    \includegraphics[width=7.5cm, angle=0]{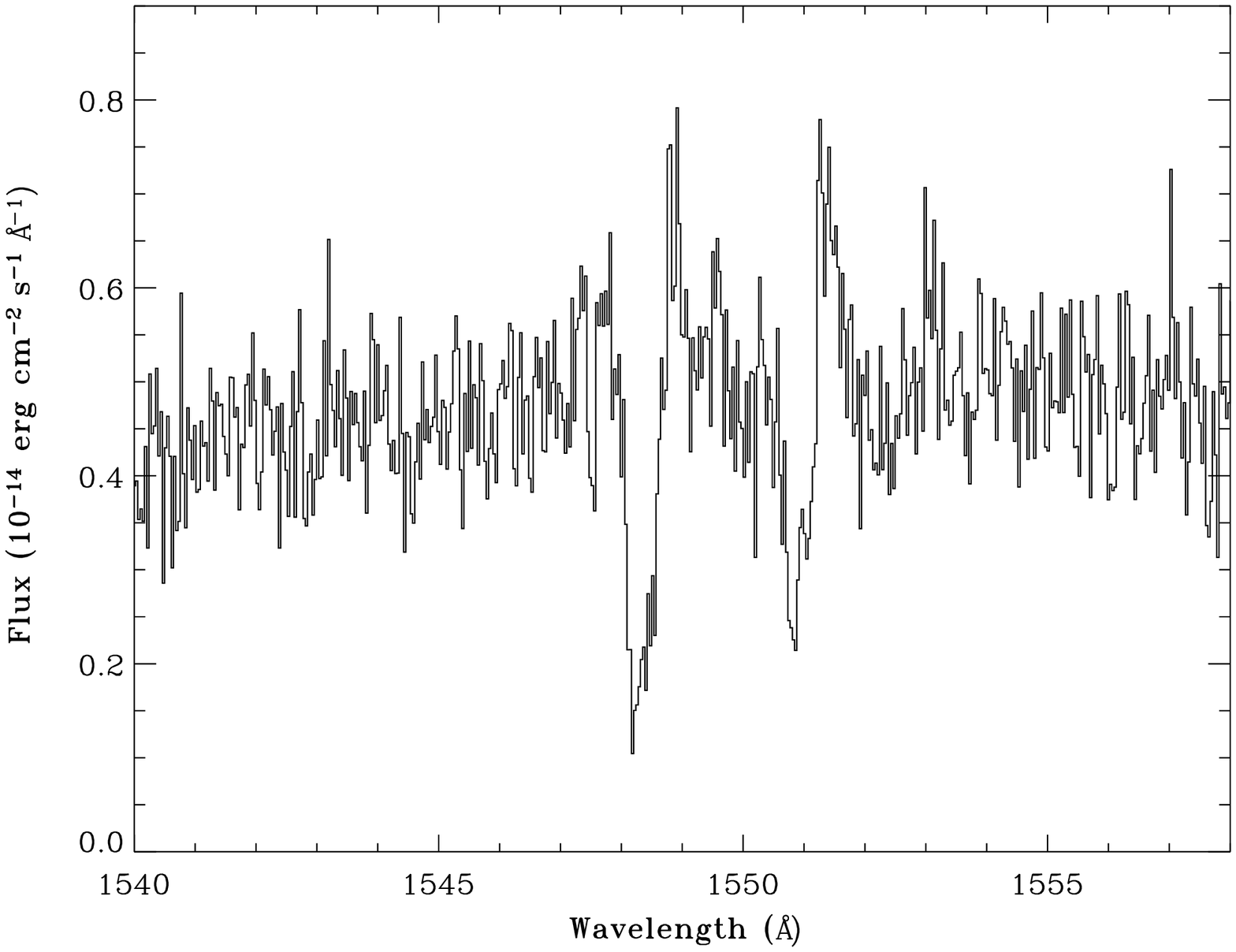}
\caption{Portion of the COS G130M (\textbf{left}) and G160M (\textbf{right}) spectrum of 
  LMC~X-3 (black histogram) showing the emission lines of the \nv\ $\lambda
  $1238.8,1242.8 doublet and the \civ\ $\lambda$
  1548.2,1550.8 doublet. The \nv\ and \civ\ doublets are both apparent in emission as 
  well as absorption. Although these features seem reminiscent of
  P-Cygni profiles, it is more likely that the absorption features are due to the absorption 
  by the foreground ISM and the emission lines are from LMC X-3. See Section 3.3 for more 
  detail. In the left panel, the smooth black curves show the
  model of the underlying LMC X-3 flux used to extract the \nv\
  emission and absorption characteristics. The red dotted curves show only
  the fitted \nv\ emission lines, and the green curves show the total
  \nv\ fits (absorption and emission). The COS data are binned to 10
  km~s$^{-1}$ pixels in this figure.}
\label{fig:COSNVALL}
\end{figure*}

We have recently observed LMC X-3 with COS using the high-resolution
G130M and G160M modes (Program 11642). The target was observed on 15
December 2009 with total exposure times of 16.9 ks (G130M) and 2.7
ks (G160M). The data were calibrated with the pipeline CALCOS (version
2.11b), and major fixed-pattern noise features due to grid wires
in the cross delay line detector were removed with a flat field
developed by K. France.  This flat field does not correct for all
types of fixed-pattern noise in COS spectra \citep[e.g.,][]{sahnow10},
but for the measurement of broad and bright emission lines presented
in this paper, the lower-level fixed pattern noise does not have a
significant impact on the results.

The regions of the COS G130M and G160M spectra of LMC X-3 covering the
\nv\ and \civ\ emission lines are shown in Figure~\ref{fig:COSNVALL}.
Emission in these species was previously detected from LMC X-3 by
\citet{cowley94} using low-resolution spectra obtained with the Faint
Object Spectrograph on {\it HST} as well as the {\it International
Ultraviolet Explorer}.  Our new COS data offer two advantages over
the previous observations: (1) the spectral resolution is $\approx$10
times higher, so variations in the velocity centroids can be measured
more accurately, and (2) the S/N is higher, and consequently the data
can be divided into phase bins for investigation of emission
variability vs. binary phase. As we will show below, the
\nv\ velocities are consistent with the optical orbital velocities of
LMC X-3 and are also consistent with the velocity trends exhibited by
the \ovi\ doublet.  The G160M spectrum is considerably noisier and the
\civ\ emission features are weaker, so in this paper we will focus on
the velocity and amplitude variations of the \nv\ lines and their
comparison to the \ovi\ emission.

\section{Results}
\subsection{New Ephemeris of LMC X-3\label{sec:opt_meas}}
\begin{figure}
\centering
    \includegraphics[width=6.0cm, angle=0]{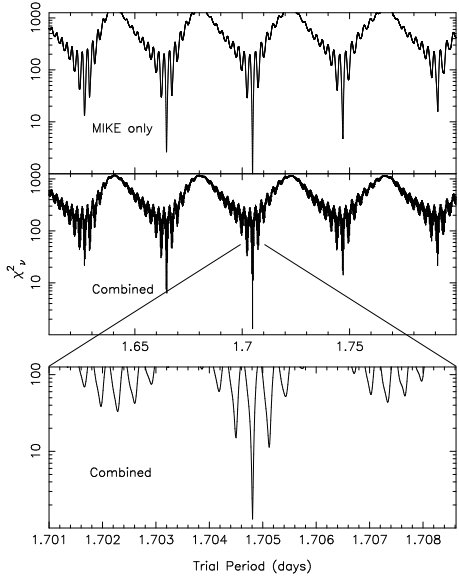} 
\caption{{\it Top:} The periodogram of LMC X-3 derived from the
Magellan+MIKE radial velocity measurements, constructed by fitting a
three-parameter sinusoid at each trial period.  The $y$-axis shows the
value of $\chi^2_{\nu}$ of the fits.  {\it Middle:} The periodogram
derived from the combination of the MIKE measurements and the Cowley
et al.\ (1983) radial velocities.  {\it Bottom:} An expanded view of
the middle panel near the minimum $\chi^2_{\nu}$. \label{period}}
\end{figure}

We have analyzed the optical observations of LMC X-3 in a manner
analogous to the recent study of LMC X-1 \citep{orosz09}. Among the 53
optical spectra that we obtained with the Magellan-Clay telescope, we
exclude seven observations from our analysis due to their relatively
large residuals in the radial velocity.\footnote{Five of the seven
outliers were obtained on the same night, so it is possible that a
subtle problem with the wavelength calibration on that particular
night compromised those measurements.} 
The remaining 46 radial velocities are then fitted to
a circular orbit model, which returns the systemic velocity $V_{\rm
0}$, the time of inferior conjunction of the optical star $T_0$ (phase
0), and the velocity semiamplitude of the secondary $K_B$, when given
the value of the orbital period $P$. We construct a
periodogram as shown in the top panel of Figure~\ref{period} using
a wide range of trial periods. The
best fit indicates $P=1.704820\pm 0.000012$ days and $T_0={\rm HJD}\,
2,454,454.9962\pm 0.0011$, where the uncertainties on the individual
measurements are scaled to give $\chi_{\nu}^{2}\approx 1$.

We also tried including the radial velocities published in C83
in order to refine our measurement. 
However, the systemic velocities obtained from the two data sets are quite 
different: 406 km $s^{-1}$ for MIKE and 310 km $s^{-1}$ for Cowley et al.
The cause for this difference is not clear, although it is not uncommon to find that 
the systemic velocity changes depending on what line is used 
(one even sees shifts when one uses different Balmer lines).
We thus first simply subtracted the respective systemic velocities from the two data 
sets before $\chi^2$ fitting sine curves as a function of trial period.
The periodogram for the combined data is shown in the middle panel of Figure
\ref{period}. The minimum value of $\chi^2_{\nu}$ occurs for
$P=1.7048089\pm 0.0000011$ days and $T_0={\rm HJD}\, 2,454,454.9964\pm
0.0011$, which are very close to the values derived from the MIKE data
alone (top panel of Figure~\ref{period}) but with much smaller 
uncertainties in P. In this paper, we use this more accurate ephemeris 
to determine the orbital phase of the UV observations of LMC X-3.
The bottom panel in
Fig. \ref{period} is the zoom-in near the minimum $\chi_{\nu}^{2}$ of
the middle panel. This panel clearly shows that all other possible
alias periods are ruled out with high confidence. 
The fitted velocity amplitude is $250.3 \pm 1.1\  km\ s^{-1}$,
comparable to the value derived by Cowley et al. (1983).
A detailed description of the optical spectroscopy and a full discussion of the
improved dynamical model used here to obtain the radial velocities
will be presented in a separate paper.
\subsection{\ovi\ Doublet Emission}
\begin{figure}
\centering
    \includegraphics[width=8.0cm, angle=0]{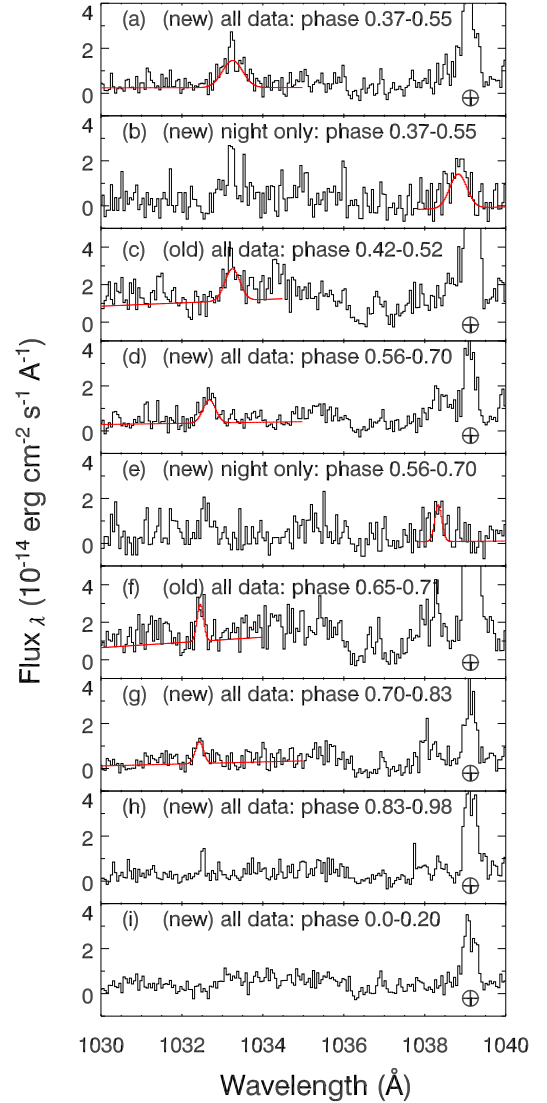}
\caption{Orbital modulation of the \ovi\ emission lines. Each panel
shows the spectrum obtained by coadding all observations in the phase
range indicated in the panel (e.g., panel a shows the data accumulated
when the XRB was between orbital phase = 0.37 and 0.55). Panels
labeled as ``new'' data are based on observations obtained in 2004,
and panels labeled as ``old'' show data from the 2001 observations.
``All data'' indicates that the spectrum includes data obtained during
both the day and night sides of the {\it FUSE} orbit and thus is
affected by terrestrial dayglow emission lines (marked with $\oplus$).
In some cases the \ovi\ $\lambda 1037.6$ transition is significantly
blended with terrestrial emission so we also show the spectrum
obtained by combining only the nighttime data (which greatly
suppresses the airglow emission features).  In some panels, Gaussian
fits to the \ovi\ emission is overplotted with a smooth red line. As
discussed in the text, the \ovi\ emission lines show regular
variations in velocity and intensity as a function of orbital
phase.\label{vE063}}
\end{figure}

The {\it FUSE} observations cover about three quarters of the XRB
orbit.  In order to measure the \ovi\ emission velocities and
intensities as a function of orbital phase, we divide the FUSE
exposures into 5 binary phase bins, which provides reasonably good
resolution of trends vs. phase while still maintaining adequate S/N in
the spectra from each bin. Fig. \ref{vE063} shows the spectra of these
5 phase bins. 

\begin{figure}
\centering
    \includegraphics[width=8.0cm, angle=0]{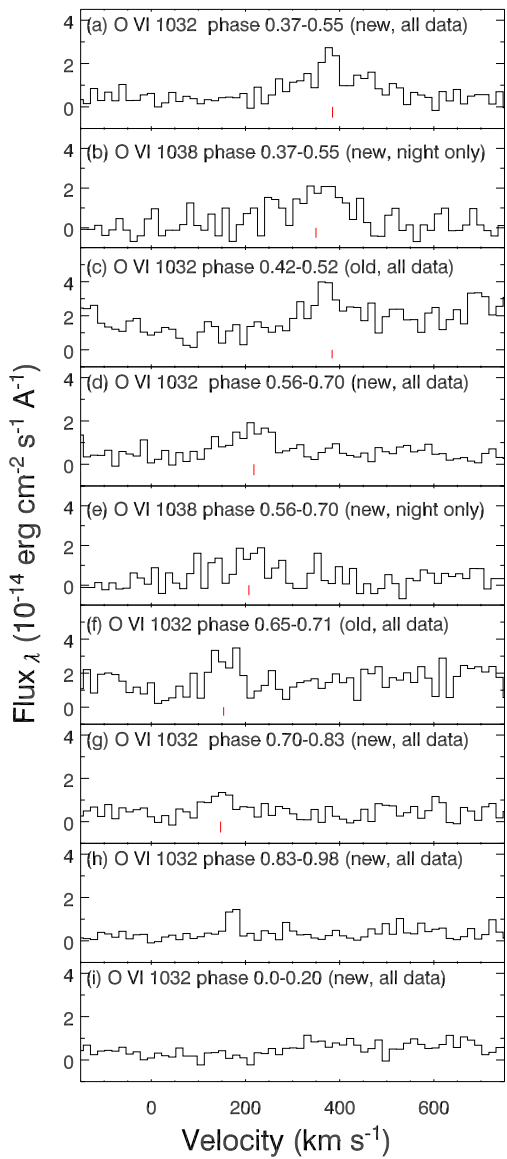}
\caption{Same data as shown in Figure~\ref{vE063} but plotted
vs. heliocentric velocity.  The zero velocities are centered at
Galactic O~VI absorption lines (1031.93 \AA \ and 1037.62 \AA),
respectively. The centroids of the emission features associated with
LMC X-3, as determined from Gaussian fits to the lines, are indicated
with red tick marks. \label{velstack}}
\end{figure}

In this and several other figures, we refer to the 2001
{\it FUSE} observations from H03 as the ``old'' data
and the later {\it FUSE} observations as the ``new'' data.  The old
data were binned in orbital phase in a similar way and are also
represented in Figure~\ref{vE063}. 
Due to the low-Earth orbit of the
{\it FUSE} satellite, many airglow emission lines from the Earth's
outer atmosphere are present in the {\it FUSE} wavelength range
\citep{feldman01}. Strong airglow emission lines are readily apparent
and easily identified (marked with $\oplus$), but weaker telluric
emission could be a source of contamination.  Most of the airglow
emission lines are excited by sunlight and are only present in data
recorded on the day side of the {\it FUSE} orbit.  Since the photons
are time tagged, it is straightforward to extract spectra using only
photons detected during the night portion of the orbit. While this
reduces the overall S/N, night-only spectra are useful for supressing
the strong airglow emission lines (which sometimes blend with -- or
mask entirely -- spectral features of interest) and for assessing
whether an emission feature is a bogus identification due to confusion
with terrestrial airglow emission. So, we also show in
Figure~\ref{vE063} some spectra extracted using night-only data.
Since the velocities of the emission lines are of particular interest,
we also show in Figure~\ref{velstack} the phase-binned {\it FUSE}
spectra plotted versus heliocentric velocity.

The spectra in Figures~\ref{vE063} and \ref{velstack} are arranged
from top to bottom according to phase (as indicated by the labels in
each panel).  From inspection of these figures, we can recognize
emission from both lines of the \ovi\ doublet that regularly shifts
in velocity as a function of the XRB orbital phase.  Interestingly,
the {\it intensity} of the \ovi\ emission also appears to change with
phase: the \ovi\ emission appears to be brightest near inferior
conjunction of the X-ray source (orbital phase $\approx$ 0.5), and it
fades and disappears into the noise at superior conjunction (phase
$\approx$ 0).  As we will show below, similar behavior is observed in
the \nv\ doublet.  To quantify this observation, we have fitted
Gaussians to the emission features, and the \ovi\ velocity centroids
and line widths indicated by the fitting exercise are summarized in
Table~\ref{o6tab}.
\begin{deluxetable*}{ccccc}
\tabletypesize{\footnotesize}
\tablewidth{0pt} \tablecolumns{5}
\tablecaption{LMC X-3 \ovi\ Emission Line Measurements\label{o6tab}}
\tablehead{\ovi\ & {\it FUSE} & Binary  & $V_{\rm O~VI}$  & FWHM \\
Transition(\AA) & Observation Date & Phase & (km s$^{-1}$) &  (km s$^{-1}$)}
\startdata
1031.93 & 2004-04 (day+night)  &0.37-0.55 & 385 $\pm$ 11 & 171 $\pm$ 26\\
1031.93 & 2001-11 (day+night)  &0.42-0.52 & 384 $\pm$ 13 & 126 $\pm$ 31\\
1037.62 & 2004-04 (night only) &0.37-0.55 & 350 $\pm$ 13 & 143 $\pm$ 32\\
1031.93 & 2004-04 (day+night)  &0.56-0.70 & 218 $\pm$ 9  & 106 $\pm$ 22\\
1037.62 & 2004-04 (night only) &0.56-0.70 & 207 $\pm$ 8  & 59  $\pm$ 16\\
1031.93 & 2001-11 (day+night)  &0.65-0.71 & 154 $\pm$ 8  & 53  $\pm$ 16\\
1031.93 & 2004-04 (day+night)  &0.70-0.83 & 147 $\pm$ 8  & 68  $\pm$ 18\\
\enddata
\end{deluxetable*}

As noted above, one concern about the identification of the
\ovi\ emission is that the features could be contaminated by
terrestrial airglow lines -- there are several molecular N$_{2}$
emission features in the vicinity of the \ovi\ $\lambda$1031.9
feature, e.g., N$_{2}$ b(1,2) and N$_{2}$ b'(9,6) \citep{feldman01},
and strong telluric emission is present near \ovi\ $\lambda$1037.6.
However, there are several indications that these emission features
we identify as \ovi\ emission lines do
not arise in the Earth's atmosphere. First, the velocity centroids of
the features regularly shift with orbital phase; this would not
occur if the lines were telluric.  Second, there are stronger N$_{2}$
features at other wavelengths in the {\it FUSE} band
\citep[see][]{feldman01}, and the absence of those stronger N$_{2}$
features indicates that the N$_{2}$ b(1,2) and N$_{2}$ b'(9,6)
emissions are negligible.  Third, in the night-only spectra the
airglow emissions in this range disappear but the candidate LMC X-3
emission lines remain apparent.  Finally, the \ovi\ trends are
corroborated by the \nv\ emission lines in the COS data, which are
clearly detected at high significance and are not affected by telluric
lines.

\subsection{\nv\ Doublet Emission}
The region of the COS G130M spectrum of LMC X-3 shown in
Figure~\ref{fig:COSNVALL} clearly shows significant emission from both
lines of \nv\ along with weaker \nv\ absorption.  Upon initial
inspection, the \nv\ features seem reminiscent of P-Cygni profiles
(the \civ\ doublet shows similar P-Cygni-like characteristics), and it
is tempting to attribute the features to a stellar wind from the
B-star companion of the LMC X-3 black hole.  While this is a possible
origin for the emission features, one must bear in mind that the
foreground highly ionized absorption from the ISM of the Milky Way (and from
the corona of the LMC) is strong in the direction of the Magellanic
Clouds \citep[e.g.,][]{wakker98,howk02}, so the \nv\ and other
highly ionized absorption is likely to arise (at least partially) in the
foreground ISM and might have nothing to do with LMC X-3 itself.  We
also note that the LMC X-3 spectral features appear to be narrower
than the P-Cygni profiles typically seen from normal B stars
\citep[cf.][]{grady87,prinja89}, and the variability of the emission
suggests an alternative origin, as we discuss below.

\begin{figure}
\centering
    \includegraphics[width=8.0cm, angle=0]{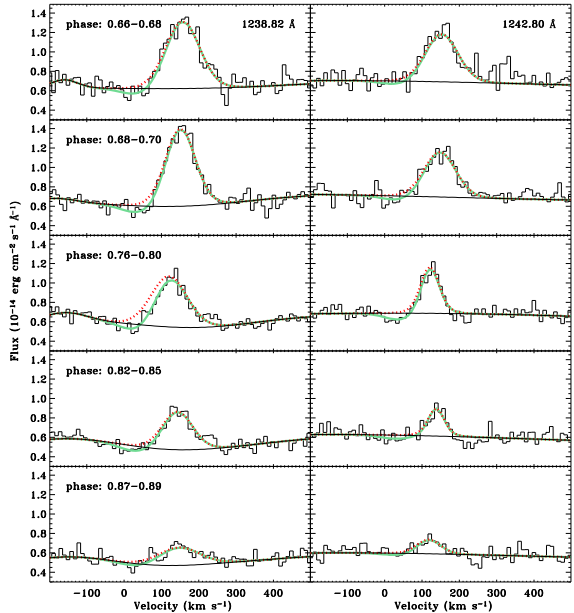}
\caption{Velocity variation of the \nv\ emission with phase, plotted
  vs. heliocentric velocity.  The zero velocities are centered at the
  rest-frame wavelengths of the doublet (the 1238.82 \AA\ and 1242.80
  \AA\ lines are shown in the left and right panels, respectively).
  The thick black lines indicate the effective continuum placement
  adopted for the \nv\ fits, the red dotted curves show the emission line
  profiles only, and the green curves show the overall fit including
  \nv\ absorption and emission. The spectra were rebinned to 10
  km~s$^{-1}$ pixels for this figure. \label{fig:COSNV}}
\end{figure}

The COS observation of the LMC X-3 covers the XRB orbit between phase
0.66 and 0.89 based on our new ephemeris. In a fashion analogous to
our treatment of the {\it FUSE} data, we divide the G130M exposures
into five phase bins and plot the velocity variation of both lines of
the \nv\ doublet with binary phase in Figure~\ref{fig:COSNV}.  We used
Gaussian and Voigt functions to model, respectively, the \nv\ emission
associated with LMC X-3 and the \nv\ absorption arising in the
foreground Milky Way gas. In each phase bin, the centroids and widths
of the Galactic \nv\ absorption lines were determined by using all the
observations (instead of only the subset data for that particular
single phase bin) since any foreground Milky Way absorption should not
change with the binary phase.  As can be seen from
Figure~\ref{fig:COSNV}, this procedure produced reasonably good
overall fits, which supports the notion that the absorption is mostly
due to foreground material unrelated to LMC X-3. It is important to
note that while the core of the COS line-spread function (LSF)
provides the expected spectral resolution, the COS LSF has broad wings
\citep{ghavamian09}.  To account for these wings, the model
profiles were convolved with the COS LSFs, at appropriate wavelengths,
from \citet{ghavamian09}.  The velocity centroids and line widths of
the \nv\ lines measured in this way are summarized in
Table~\ref{tab:COSNV}.

\begin{deluxetable*}{cccccc}
\tabletypesize{\footnotesize}
\tablecaption{LMC X-3 \nv\ Emission Line Measurements\label{tab:COSNV}}
\tablehead{\nv\ & \multicolumn{2}{c}{{\it COS} Observation} & Binary & $V_{\rm N~V}$ & FWHM \\
Transition(\AA) & Start Time(UT) & Duration(s) &  Phase &  (km s$^{-1}$) & (km s$^{-1}$)}
\startdata
1238.82 & 2009-12-15 04:07:56  & 1885 &  0.66-0.68 & $142\pm4$  & $96\pm7$ \\
1242.80 & 2009-12-15 04:07:56  & 1885 &  0.66-0.68 & $140\pm3$  & $101\pm8$ \\
1238.82 & 2009-12-15 04:47:49  & 3014 &  0.68-0.70 & $134\pm4$  & $94\pm7$ \\ 
1242.80 & 2009-12-15 04:47:49  & 3014 &  0.68-0.70 & $131\pm3$  & $113\pm8$ \\
1238.82 & 2009-12-15 07:57:18  & 5411 &  0.76-0.80 & $125\pm4$  & $87\pm10$ \\
1242.80 & 2009-12-15 07:57:18  & 5411 &  0.76-0.80 & $126\pm3$  & $45\pm5$ \\
1238.82 & 2009-12-15 10:40:19  & 3270 &  0.82-0.85 & $145\pm6$  & $70\pm11$ \\
1242.80 & 2009-12-15 10:40:19  & 3270 &  0.82-0.85 & $140\pm5$  & $44\pm8$ \\
1238.82 & 2009-12-15 12:22:19  & 3270 &  0.87-0.89 & $136\pm11$ & $113\pm23$ \\
1242.80 & 2009-12-15 12:22:19  & 3270 &  0.87-0.89 & $124\pm10$ & $46\pm9$ \\
\enddata
\end{deluxetable*}

\begin{figure*}
\centering
    \includegraphics[width=16.0cm, angle=0]{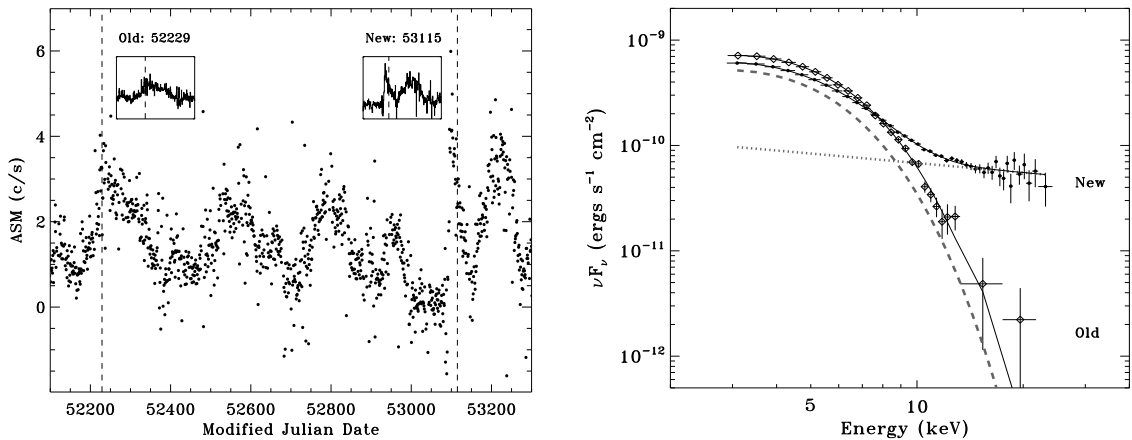}
\caption{{\it Left:} X-ray light curve of LMC X-3 from the {\it RXTE}
All-Sky Monitor with the dates of the ``old'' (2001) and ``new''
(2004) {\it FUSE} observations marked (dashed lines). The two insets show 
the regions in the vicinity of the two epochs with a time span of 350 days.
{\it Right: RXTE Proportional Counter
Array} X-ray spectral energy distribution of LMC X-3 obtained near the times of the two {\it FUSE}
observations. The solid lines are the best-fit models to the data (see Table~\ref{tab:rxte} for model details). The grey dashed and dotted lines indicate the disk and power-law components used to fit
the new observation, respectively. \label{asmsed}}
\end{figure*}
As we can see from Figure~\ref{fig:COSNV} and Table~\ref{tab:COSNV},
the velocity centroids of the \nv\ emission lines do not change much
in the five phase bins extracted from the COS data set.  However, the
COS observations were randomly scheduled with respect to the XRB phase
(i.e., no scheduling constraints were placed on the COS program in
order to observe the system at some particular phase), and as we will
show in \S \ref{sec:disc}, the COS observations happened to be obtained
near an extremum of the XRB radial velocity curve when the velocities
are not expected to change much.  Moreover, the \nv\ velocity
centroids are consistent with the \ovi\ velocities recorded in roughly
the same phase (see below).  Perhaps more interestingly, the \nv\ {\it
intensities} change with phase in the same way as the \ovi: the
\nv\ emission intensity decreases steadily going from phase $\approx
0.65$ to phase $\approx$ 0.9 (see Figure ~\ref{fig:COSNV}).  This is
surely an important clue about the origin of the \nv\ and
\ovi\ emission, and we will return to this in \S \ref{sec:disc}.

\subsection{X-ray Observations\label{sec:xrayobs}}
Figure~\ref{asmsed} compares the X-ray characteristics of LMC X-3
during the {\it FUSE} observations in 2001 and 2004. Although LMC X-3
usually occupies a soft state in which the X-ray spectrum is dominated
by emission from the accretion disk, it was in a faint and transitional state
near the time of the 2004 {\it FUSE} observations \citep[this is
indicated by the presence of a strong power-law component in the
observed spectrum; see][]{cui02}.  As shown in Figure~\ref{asmsed},
LMC X-3 appeared to be nearly as bright in X-rays during the 2004 {\it
FUSE} observations as it was during the 2001 observations of
H03. This is not accidental; the 2004 {\it FUSE}
observations were part of a coordinated {\it Chandra/FUSE}
target-of-opportunity program that was triggered when the {\it RXTE}
ASM showed that LMC X-3 was entering a bright state.\footnote{This was
done to maximize the S/N of the {\it Chandra} grating spectroscopy,
which was originally initiated for the detection of X-ray absorption
lines. See \citet{wang05} for further details.}  However, as can be
seen from the left panel of Figure~\ref{asmsed}, the 2004 increase in
X-ray brightness was associated
with a short-lived X-ray flare, whereas
the 2001 data were recorded when LMC X-3 was undergoing one of its more
gradual (and more typical) upward modulations in X-ray brightness.
Thus, while the X-ray brightness was comparable in 2001 and 2004, it
is likely that the object was in a different type of X-ray state.  To
further compare the X-ray characteristics of the XRB during the {\it
FUSE} observations, the right panel of Figure~\ref{asmsed} shows
spectra from pointed {\it RXTE} observations obtained very close to
the times of the {\it FUSE} studies. From this comparison we see that
LMC X-3 was in the usual soft state around the time of the 2001
observations and the associated spectrum is well fitted with a multicolor disk model. However, 
LMC X-3 was in a significantly harder spectral state 
when the 2004 {\it FUSE} data were obtained. The spectrum for the latter observation requires
a second power-law component aside the disk component. The best-fit parameters from the 
two models are summarized in Table~\ref{tab:rxte}. 
\begin{deluxetable*}{cccccc}
\tablewidth{0pt}
\tablecaption{Best-fit Parameters for RXTE Spectra of LMC X-3$^a$
\label{tab:rxte}}
\tablehead{ &  & \multicolumn{2}{c}{\underline{Disk Component$^b$}}
         & \multicolumn{2}{c}{\underline{Power-law Component$^c$}} \\
      Observation   & $N_{\rm H}$ & $kT_{\rm dbb}$ &  & &  \\
    Time  & ($10^{20}$ cm$^{-2}$) & (keV) & $N_{\rm dbb}$ & $\Gamma$ &
$N_{\rm po}$ }
\startdata
2001 & 4.74 & $1.25_{-0.01}^{+0.01}$ & $27.85_{-0.94}^{+0.98}$  &
---
& --- \\
2004 & 4.74 & $1.19_{-0.01}^{+0.01}$ & $24.45_{-1.05}^{+1.03}$ &
$2.30_{-0.16}^{+0.15}$ & $0.085_{-0.028}^{+0.039}$ \\
\enddata
\tablenotetext{a}{The uncertainties shown represent 90\% confidence
intervals.}
\tablenotetext{b}{$T_{\rm dbb}$ and $N_{\rm dbb}$ are the temperature at
inner disk radius and
the normalization (dimensionless) in the multi-black body model ({\it
diskbb} in XSPEC), respectively.}
\tablenotetext{c}{$\Gamma$ and $N_{\rm po}$ are the photon index of
power law and the normalization (photons keV$^{-1}$ cm$^{-2}$ s$^{-1}$
at 1 keV) in the power law model ({\it powerlaw} in XSPEC),
respectively.}
\end{deluxetable*}

\section{Discussion: Nature of the UV Line Emission\label{sec:disc}}

Where do the ultraviolet emission lines originate?  Emission from
\ovi\ arising in interstellar plasma has been frequently detected
\citep[e.g.,][]{dixon06}, but strong \nv\ emission lines like those
shown in Figure~\ref{fig:COSNVALL} are never seen from interstellar
clouds, so it is most likely these emission lines are intrinsic to the
LMC X-3 system. Moreover, lines from interstellar gas should be
stationary, but the LMC X-3 emission-line velocities seem to follow
the velocity curve of the XRB.  

\begin{figure}
\centering
    \includegraphics[width=8.0cm, angle=0]{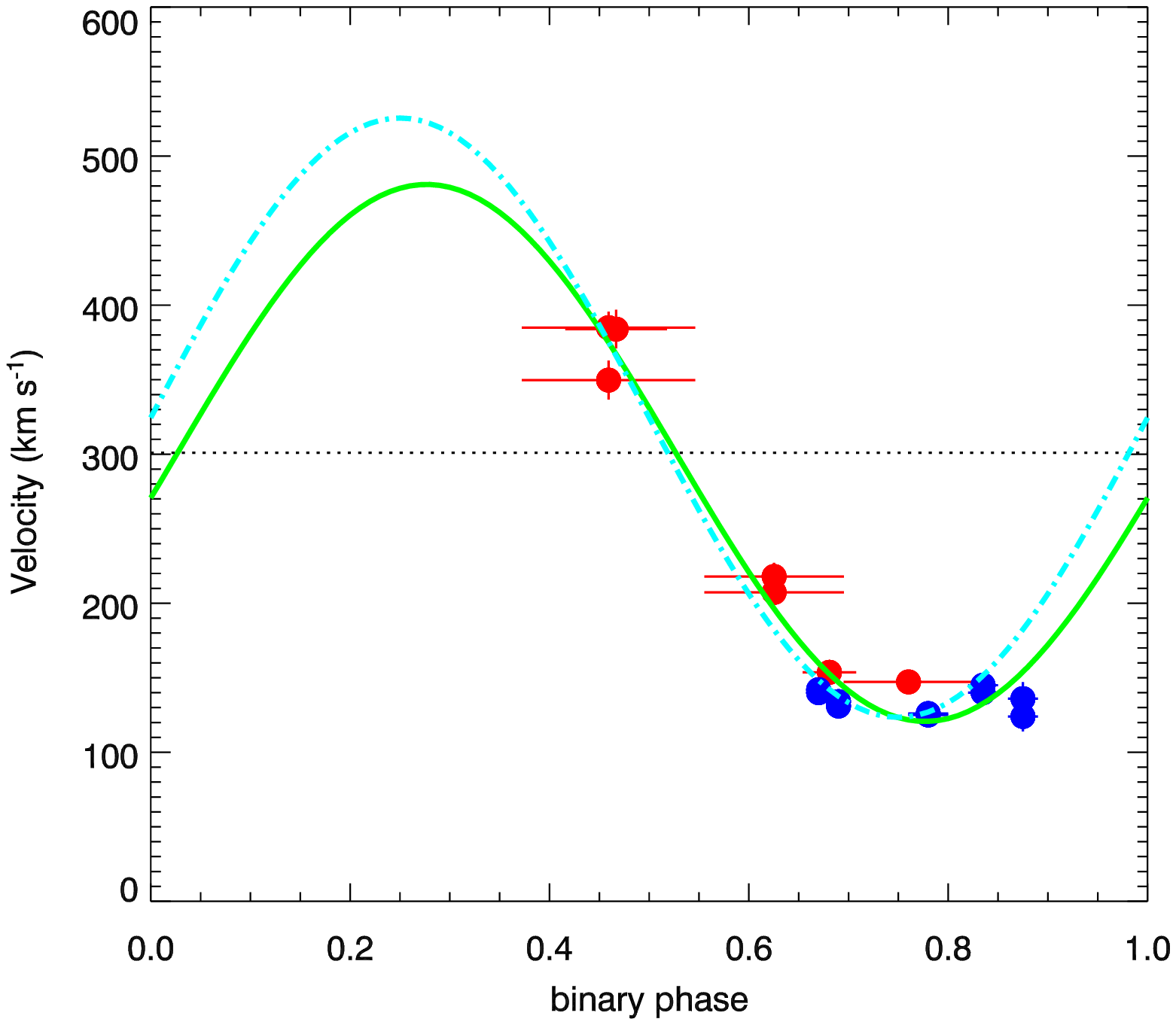}
\caption{Velocity variation with binary phase for the \ovi\ 
(red filled circles) and \nv\ (blue filled circles) emission lines. 
No measurements are available for the \ovi\ emission lines in the phase 
range 0.83-0.20 since the emission is too weak to be detected near 
the X-ray source superior conjunction.
Horizontal and vertical bars on each point indicate the phase coverage and 
$\pm 1\sigma$ velocity uncertainties associated with each measurement, respectively.
A sinusoidal fit to the UV data is presented as the 
green curve. The cyan dashed-dotted line shows a fit to the UV data in which
the sinusoid is forced to be in phase with the optical curve but the semiamplitude
is allowed to vary. The systematic velocity of LMC X-3 from the fit to the UV data 
(301 km $s^{-1}$) is marked as the dotted black line. \label{fitcurve}}.
\end{figure}

To show the UV emission-line velocity trends more quantitatively, we
plot the velocity centroids of the \ovi\ and \nv\ emission lines as a
function of the binary phase (derived from the new ephemeris) in
Figure~\ref{fitcurve}. In this figure, the \ovi\ and \nv\ emission
velocities are plotted with red and blue filled circles, respectively;
the horizontal bars on each point indicate the phase range associated
with each measurement, and the vertical bars represent the formal $\pm
1\sigma$ velocity uncertainties. 
When we fit a sinusoidal
curve to the UV data (green curve), we find an orbital semiamplitude
of $180\pm 1\ km\ s^{-1}$ and a phase deviation (compared to the optical
data) of $0.027\pm 0.004$. The systemic velocity of LMC X-3 inferred from the
fit to the UV measurements is 301 km $s^{-1}$, comparable to the
systemic velocity ($310 \ \pm$ 7 km s$^{-1}$) obtained in C83, assuming a circular orbit.
The cyan dashed-dotted line shows a fit to the UV velocities in which the sinusoid 
is forced to be in phase with the optical curve but the semiamplitude is allowed to vary; 
we see that this fit is comparably good. Therefore, the UV emission velocities generally follow
the trend of the optical velocities in the orbital phase, but the UV velocity semiamplitude
is lower than the optical semiamplitude ($250.3 \pm 1.1 \  km\ s^{-1}$).

If the \ovi\ and \nv\ emission lines arise in the atmosphere of the B
star, they should follow the optical orbital velocities of the star.
For a circular orbit, which is most likely for LMC X-3 since the 1.7 d
orbit should quickly circularize, the radial velocity curve should be
a standard sinusoid. This behavior is approximately seen in the UV
emission velocities shown in Figure~\ref{fitcurve}.
It is possible that there are systematic
uncertainties in the UV velocities (e.g., due to line shapes that are
more complex than our assumed single Gaussian plus interstellar
absorption) that are not adequately reflected by the formal error bars
from the single-Gaussian fits, but the velocities are sufficiently
well-constrained to show that they follow the general velocity trend
of the optical B-star measurements and that the UV semiamplitude is clearly
different from the optical semiamplitude.

\begin{figure}
\centering
    \includegraphics[width=8.0cm, angle=0]{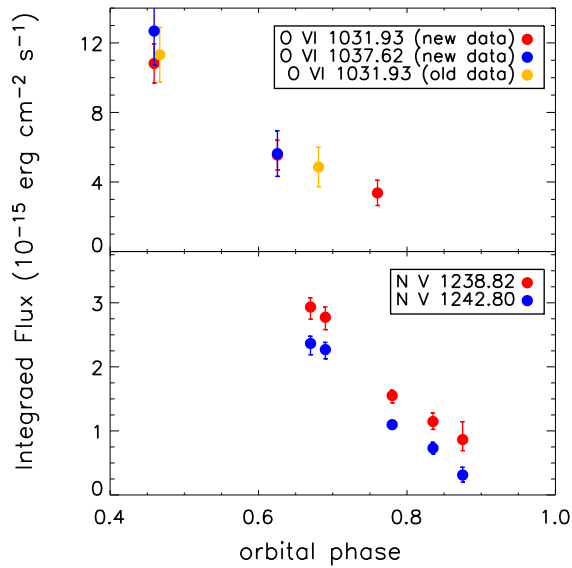}
\caption{ Variability of the integrated flux of \ovi\ and \nv\ emission as a function of 
orbital phase. \label{fluxphase}}.
\end{figure}

The similar (but not identical) behavior of the UV vs. optical
velocities could naturally occur if one side of the B star is illuminated 
and hence heated by the X-ray/UV
continuum emission originating near the black hole.
In this case, the UV and B-star orbital
velocities would generally follow each other, but rotation
of the B star could cause the UV emission to have a somewhat different
semiamplitude. Cowley et al. (C83) report that the B-star has $v$ sin
$i$ = $130\pm 20$ km s$^{-1}$. A hot spot on the B star that
remains pointing in the direction of the black hole would only be
rotating toward or away from the observer depending on where the star
is in the course of its orbit, and since the velocity centroid would
be some sort of weighted mean of the portion of the spot visible to
the observer, the effective component of the rotation velocity along
the line of sight could be less than the full $v$ sin $i$ value.  To
test this hypothesis, it would be helpful to obtain observations of
LMC X-3 covering the phase = 0.2 $-$ 0.5 portion of the orbit.  It
would also be helpful to carry out detailed modeling of this scenario,
but such modeling is beyond the scope of this paper.

However, we note that the {\it intensity} variations of the \ovi\ and
\nv\ emission lines are broadly consistent with this idea.
Figure~\ref{fluxphase} shows the variations of the integrated flux in
the \ovi\ and \nv\ emission vs. orbital phase.  The gradual decrease
of the integrated \ovi\ and \nv\ flux occurs over binary phases when
the accreting black hole moves from inferior conjunction to quadrature
to superior conjunction.  At superior conjunction, the \ovi\ emission
is too weak to be reliably measured, and the \nv\ emission is greatly
suppressed (see Figures \ref{velstack} and \ref{fig:COSNV}).  During
this time period, the hemisphere of the B star directly facing the
black hole would move from fully illuminated to partially covered to
completely occulted. The observed UV emission intensity variations
appear to be consistent with this hypothesis.

A similar hypothesis has been proposed to explain the similar \nv\ 
emission lines and other emission features detected in the UV spectra of 
a low-mass X-ray binary 
Cygnus X-2 \citep{vrtilek03}. By measuring the radial velocities
of the emission lines and comparing the observed line
profiles with the predictions from models of line emission from an 
X-ray-heated accretion disk corona, Vrtilek et al. (2003) suggest that 
most of the line emission detected in the UV spectra of Cygnus X-2 
is from the illuminated surface of the 
companion star. For LMC X-3, if the idea is supported by future work, this could provide
an important constraint on the inclination of this X-ray binary (and
thus the mass of the black hole) as well as insight on how black holes
affect their surroundings, and the properties of their companions in particular.

Our interpretation of the detected variable \ovi\ and \nv\ emission from the X-ray binary 
LMC X-3 seems promising, but it is far from being conclusive. There exists considerable uncertainties in decomposition of the \nv\ emission feature (\S 3.3). There could be weak P-Cygni profiles in the N
V and C IV regions.  For the absorption part of the P-Cygni profile,
the importance and shape of the wind contribution versus the contribution from the foreground ISM is difficult to distinguish and accurately measure. 
We also can not fully exclude contributions from other plausible components of the LMC X-3 system, such 
as stellar and/or accretion disk winds as detected or predicted in other systems 
\citep{boroson07,raymond93}. Our hypothesis requires further observational and theoretical
testing.

\acknowledgements We would like to thank John Raymond, Saku Vrtilek and Jifeng Liu 
for their very helpful comments on an early version of the paper. We thank the referee for thoughtful suggestions.  
The data presented in this paper were obtained for
the {\it FUSE} GO program E063 and the {\it HST} GO program 11642.
These programs were funded by NASA through grants NNG04GJB83G, HST-GO-11642.01-A, 
and HST-GO-11642.02-A, respectively.

\end{document}